\documentclass[fleqn]{annmod}
\usepackage{epsf}
\begin{document}
\newcommand{\volume}{8}              
\newcommand{\xyear}{1999}            
\newcommand{\issue}{5}               
\newcommand{\recdate}{29 July 1999}  
\newcommand{\revdate}{dd.mm.yyyy}    
\newcommand{\revnum}{0}              
\newcommand{\accdate}{dd.mm.yyyy}    
\newcommand{\coeditor}{ue}           
\newcommand{\firstpage}{1}           
\newcommand{\lastpage}{4}            
\setcounter{page}{\firstpage}        
\newcommand{\keywords}{disordered systems, metal-insulator transitions, 
                       quantum chaos} 
\newcommand{\PACS}{71.23.-k, 71.30.+h, 05.45.Pq}
\newcommand{\shorttitle}{D.\ Weinmann {\it et al.}, 
 Disordered Systems in Phase Space} 
\title{Disordered Systems in Phase Space}
\author{Dietmar Weinmann$^{1}$, Sigmund Kohler$^{1,2}$, 
        Gert-Ludwig Ingold$^{1}$, \\ and Peter H\"anggi$^{1}$} 
\newcommand{\address}{$^{1}$Institut f\"ur Physik, Universit\"at Augsburg, 
             86135 Augsburg, Germany\\
             $^{2}$Faculdad de Ciencias, Universidad Autonoma de Madrid, 
             28049 Madrid, Spain}
\newcommand{\email}{\tt Gert.Ingold@Physik.Uni-Augsburg.DE} 
\maketitle
\begin{abstract}
As a function of the disorder strength in a mesoscopic system, the electron
dynamics crosses over from the ballistic through the diffusive towards the 
localized regime. The ballistic and the localized situation correspond to 
integrable or regular behavior while diffusive conductors correspond to 
chaotic behavior. 
The chaotic or regular character of single wave functions can be 
inferred from phase space concepts like the Husimi distribution and the
Wehrl entropy. These quantities provide useful information about the 
structure of states in disordered systems. 
We investigate the phase space structure of one dimensional (1d) and 2d 
disordered systems within the Anderson model. The Wehrl entropy of the 
eigenstates allows to detect the crossover between the ballistic, diffusive 
and localized regime.
\end{abstract}

\vspace{0.5cm}
The presence of a disordered potential in mesoscopic systems strongly 
affects the structure of the quantum states and thereby the electronic 
properties of the sample. The disorder results in a finite elastic mean free 
path $l$ and can lead to exponential localization of the wave functions with 
localization length $\xi$ \cite{anderson}. 
The electronic properties of the sample can be classified according to the 
ratios $l/L$ and $\xi/L$ of these length scales to the system size $L$.
When $l > L$, the dynamics is ballistic, the wave functions being 
close to the plain waves with fixed momentum value found in clean systems. 
In contrast, the regime $\xi < L$ is called localized since the wave functions
are restricted to a finite domain in real space. This implies delocalization 
in momentum space. 
A crossover between wave function structures localized in momentum 
space and localized in real space occurs as a function of the disorder 
strength. In the intermediate regime, when $l < L < \xi$, the dynamics of the 
particles in the disordered potential is diffusive. The particles are 
scattered several times while traversing the system, but the localization of 
the wave function is not relevant on the scale $L$, leading to 
a complex structure of the wave function.  
The crossover between the different regimes is accompanied by a 
change of not only the structure of wave functions, but also the energy level 
statistics. This suggests that disordered systems in the 
diffusive regime exhibit chaos in contrast to the ballistic and the 
localized regime which are both classified as integrable.   
  
In order to investigate the changes of the structure of the wave functions
throughout all regimes, it is desirable to use a concept which allows to 
detect both, changes in real space and in momentum space structure.
A quantum state $|\psi\rangle$ is completely 
determined by its wave function in real space representation 
$\psi(x) = \langle x | \psi \rangle $ or by its momentum space representation.
It is nevertheless possible to construct quantities which depend on both, 
position and momentum $p=\hbar k$, and to study the structure of the 
wave functions in phase space $(x,k)$. The Husimi density characterizing a 
given state $|\psi\rangle $,
\begin{equation}
 \rho_{\rm H}(x_0,k_0) = \left|\langle x_0,k_0 | \psi\rangle \right|^2\, ,
\end{equation}
is given by the projection of $|\psi\rangle $ on minimal uncertainty states 
$|x_0,k_0\rangle$. 
We use Gaussian wave packets with variance $\sigma^2$, centered around 
position $x_0$ and wave number $k_0$. In real space representation, 
these states are given by 
\begin{equation}\label{coherent}
 \langle x|x_0,k_0\rangle = 
  \left(\frac{1}{2\pi\sigma^2} \right)^{1/4} 
  \exp\left(-\frac{(x-x_0)^2}{4\sigma^2}+{\rm i} k_0 x\right)\, .
\end{equation} 
This definition of $\rho_{\rm H}$ yields the normalization
$\int\frac{{\rm d}x{\rm d}k}{2\pi}\rho_{\rm H}(x,k)=1$.
The resulting Husimi-distribution can be used to 
visualize the phase space structure of quantum states. Its analysis
allows to detect structures in wave functions which 
correspond to the classical dynamics in phase space \cite{takahashi}. 
Since $\rho_{\rm H}$ is always nonnegative, the Wehrl entropy
\begin{equation}
S_{\rm H}= - \int \frac{{\rm d}x{\rm d}k}{2\pi} \,\rho_{\rm H}(x,k) 
\ln \left[\rho_{\rm H}(x,k)\right]
\end{equation}
can be defined \cite{wehrl,mirbach}. $S_{\rm H}$ is a measure of the 
phase space volume occupied by the quantum state.
It has been shown for the driven rotor that the Wehrl entropy 
of individual states is connected to the energy level statistics. 
Both quantities can be used to distinguish between the chaotic and the 
integrable regime \cite{gorin}. 
A very similar system, the kicked rotor, can be mapped onto the Anderson 
model \cite{fishman}, suggesting that the Wehrl entropy is a useful quantity 
for the characterization of the eigenstates of the Anderson model. 

The Anderson Hamiltonian \cite{anderson} describes a particle
on a lattice. Its 1d version reads
\begin{equation}\label{anderson}
H_{\rm A}=-t \sum_n
\left( |n\rangle\langle n+1| + |n+1\rangle\langle n | \right)
+ W \sum_n v_n \, |n\rangle\langle n| 
\end{equation}
with Wannier states $|n\rangle$ localized at sites $n$. In units of the 
lattice spacing, the position of a site is $x_n=n$.
The hopping matrix elements $t=1$ between neighboring sites define the 
energy scale. The on-site disorder $v_n$ is drawn from a box distribution 
inside the interval $[-1/2;1/2]$, and $W$ denotes the disorder strength. 

\begin{figure}[tbh]
\centerline{\epsfxsize=13cm\epsffile{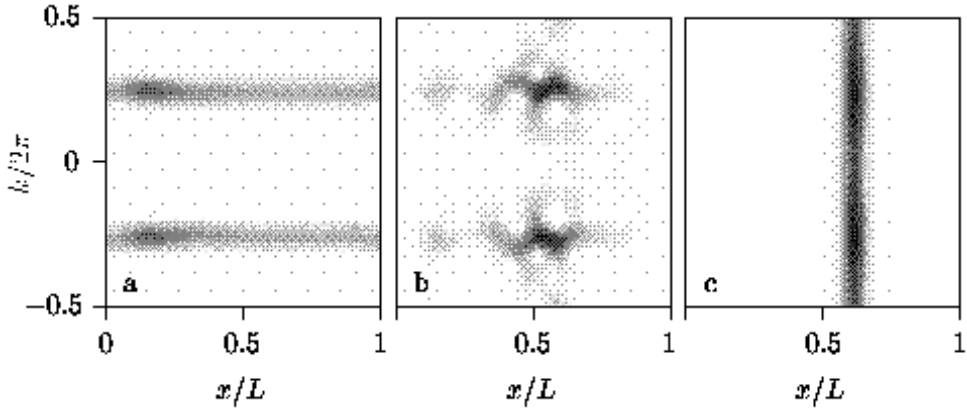}}
 \caption{\label{husimi1d} Husimi distribution of an eigenstate of the
 1d Anderson model of length $L=100$ at $W=0.5$ (a), 3 (b), and 300 (c).}
\end{figure}

We have evaluated the Husimi distributions for 1d rings 
containing $L=100$ sites with periodic boundary conditions 
$|n\rangle \equiv |n+L\rangle$.
In order to adapt the minimum uncertainty states to the lattice structure and 
the ring geometry of our model, we use the form (\ref{coherent}) in the 
interval $x_0-L/2 \le x \le x_0+L/2$ such that the tails of the Gaussian in 
real space are cut opposite to the position of the maximum $x_0$. The lattice 
determines the possible values of the wave numbers $k_0=2\pi j/L$, with 
$j=\{-L/2+1,-L/2+2,\dots,L/2\}$ ($L$ even) for the first Brillouin zone. The 
phase space $(x,k)$ is then represented by a $L \times L$ lattice.
The choice $\sigma=\Delta x =\sqrt{L/4\pi}$ in (\ref{coherent}) ensures that 
the minimum uncertainty states with $\Delta x \Delta k = 1/2$ are extended in 
$x$ and in $k$ direction over the same number of allowed discrete values. 
 
The eigenstates of (\ref{anderson}) at $W=0$ are plain waves. 
This means complete delocalization in real space and maximum 
localization in momentum space. Obeying the symmetry 
$\rho_{\rm H}(x,k)=\rho_{\rm H}(x,-k)$, the corresponding Husimi density, 
presented in Fig.\ \ref{husimi1d}a, reflects this structure by exhibiting 
two lines parallel to the $x$-axis. 
In the limit of very strong disorder, the 
wave functions are completely localized in real 
space but delocalized in momentum space. Then, the Husimi density
(Fig.\ \ref{husimi1d}c)
is given by a line parallel to the $k$-axis. These phase space structures 
are typical for the ballistic ($l \gg L$) and the localized ($\xi \ll L$) 
regime, respectively. Since $l\approx \xi$ in 1d, no diffusive 
regime with chaotic wave functions appears.  
In the crossover regime, when $L$ equals a few $\xi$, 
the Husimi density is subject to strong fluctuations. The typical situation is 
intermediate between the scenarios described above, and the density is more or 
less extended in both variables, position and wave number, as can be seen in 
Fig.\ \ref{husimi1d}b. 

\begin{figure}[tbh]
\centerline{\epsfxsize=6cm\epsffile{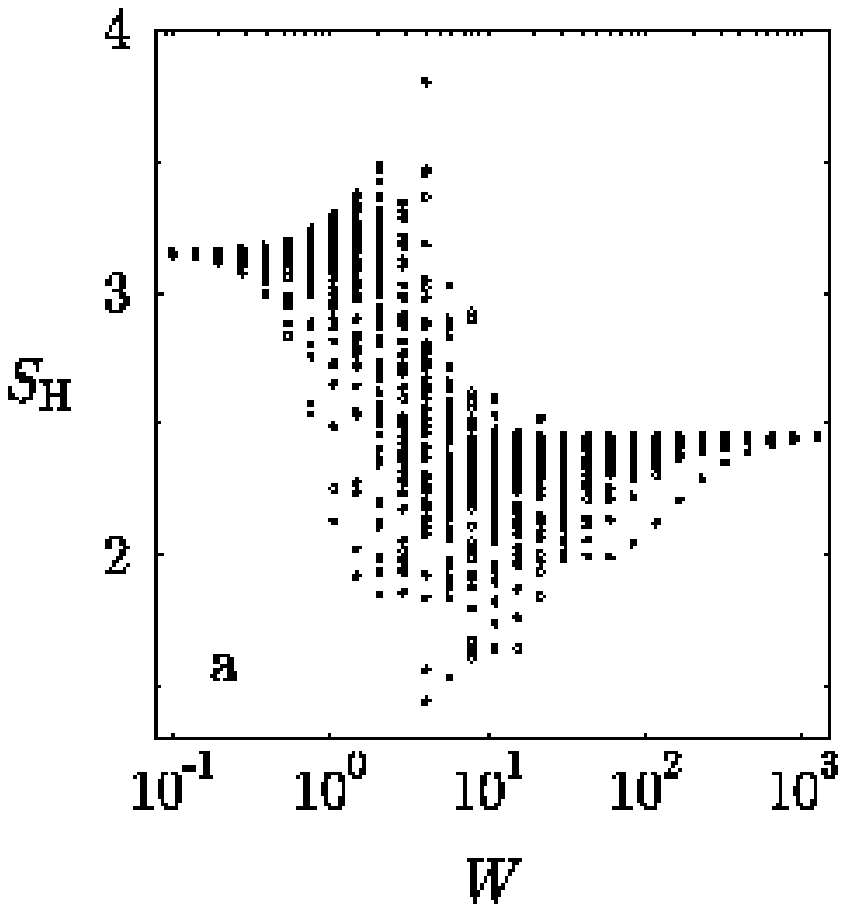}
 \hfill \epsfxsize=6cm\epsffile{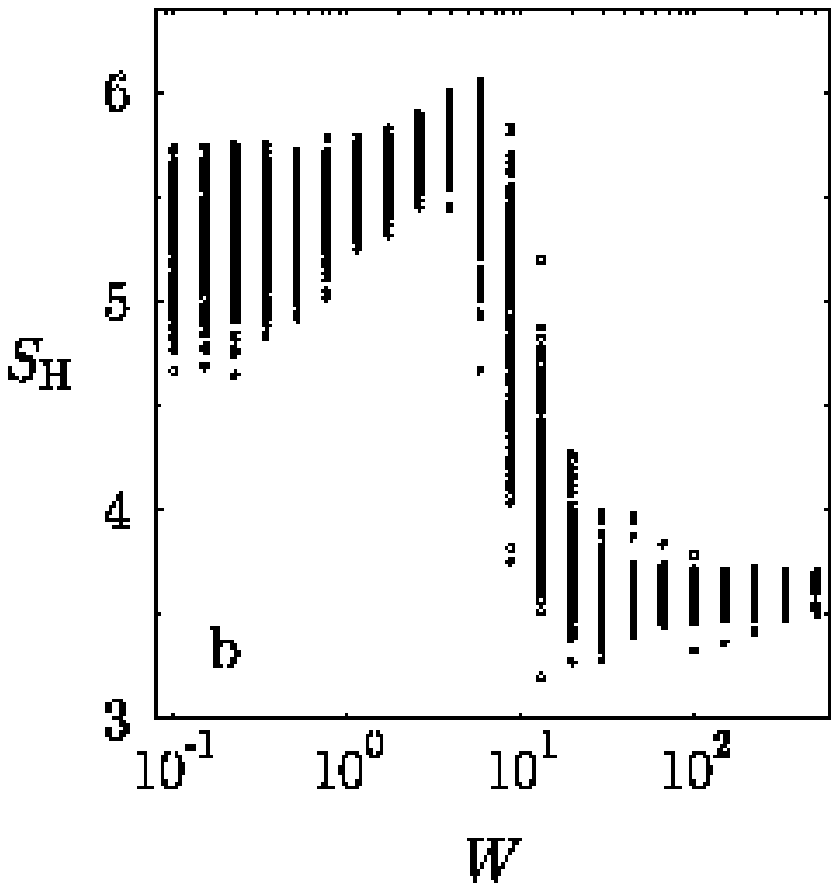}}
 \caption{\label{entropies} a: Wehrl entropies for the states of the 
 1d Anderson model of length $L=100$. Data for 60 states around the band 
 center are shown. b: Wehrl entropies for states 180-359 on a 2d 
 lattice of $30\times 30$ sites.}
\end{figure}

Fig.\ \ref{entropies}a shows the Wehrl entropies of 60 eigenstates 
around the band center for a ring of size 
$L=100$ as a function of the disorder.
While the average of the Wehrl entropy in 1d decreases when 
the disorder is increased ($\rho_{\rm H}$ exhibits two stripes when 
$W\rightarrow 0$, while for $W\rightarrow \infty$, only one is present, 
see Fig.\ \ref{husimi1d}), its variance clearly exhibits a maximum in the 
crossover region around $W\approx 3$. The entropies for states at the band 
edge, where the two stripes at $k$ and $-k$ in the ballistic regime overlap 
are not shown, leading to well-defined values in both limits $W\rightarrow 0$ 
and $W\rightarrow \infty$. 

Using a straightforward extension of the above concepts to 2d systems, we have 
also investigated the phase space structure of the eigenstates of the 2d 
Anderson model. The phase space $(x_1,x_2;k_1,k_2)$ is now four-dimensional 
and the corresponding minimum uncertainty states are given by products of two 
terms of the form (\ref{coherent}).
While it becomes difficult to visualize the Husimi density, the Wehrl 
entropy, now being obtained from a four-dimensional integral, still provides
a measure for the phase space occupation of the state. In 
Fig.\ \ref{entropies}b, we plot the entropies for a few states below the band 
center in order to avoid artifacts arising at low disorder from the 
singularity of the density of states. 
In contrast to the 1d situation, the
localization length is always larger than the mean free path in 2d, 
and the diffusive regime appears when $l<L<\xi$ (for our case $L=30$, using 
data from \cite{kmk}, this 
corresponds to the disorder interval between $W\approx 1.5$ and $W\approx 6$). 
This situation is accompanied by an enhancement of the Wehrl entropy, 
confirming the expected chaotic character of the wave functions. 
The ballistic and localized limits remain regular, being characterized by 
smaller values of the Wehrl entropy. 
While all entropies converge towards the same value when $W\rightarrow\infty$, 
a certain variance of entropies remains in the limit $W\rightarrow 0$. 
In contrast to the 1d case, this cannot be avoided because
it is not possible to exclude the few states with overlapping stripes in the 
ballistic regime by selecting a certain energy 
interval. 

In conclusion, we have shown that phase space concepts indeed provide useful 
information about the states of disordered systems. In particular, they allow 
to distinguish the different transport regimes.
It will be interesting to use this new method for detailed investigations of 
the Anderson metal-insulator transition, as well as for correlations in 
many-body problems.

\vspace{2mm}
{\small
S.K.\ thanks the European Union for financial support within the
TMR network ``Phase coherent dynamics of hybrid nanostructures''.}


\end{document}